\documentclass[11pt]{article}
\usepackage{hanoi,epsfig}

\usepackage{graphicx,amsmath,amssymb}

\renewcommand{\leq}{\leqslant}

\newcommand{\ket}[1]{|\kern.3ex#1\kern.3ex\rangle}
\newcommand{\bra}[1]{\langle\kern.3ex #1 \kern.3ex|}

\newcommand{\smean}[1]{\langle #1 \rangle} 

\newcommand{\EXP}[1]{{\mbox{\large e}}^{#1}}         

\def\I{{\rm i}}                  
\def\D{{\rm d}}                  

\begin{document}
\title{MAGNETOCONDUCTANCE OSCILLATIONS IN METALLIC RINGS \&\\[0.05cm]
  DECOHERENCE DUE TO ELECTRON-ELECTRON INTERACTION}

\author{ Christophe Texier$^{\dagger,\ddagger}$ and Gilles Montambaux$^\ddagger$ }

\address{$^\dagger$Laboratoire de Physique Th\'eorique et Mod\`eles
  Statistiques, UMR 8626, CNRS,Universit\'e Paris-Sud, F-91405 Orsay Cedex, France.\\
$^\ddagger$Laboratoire de Physique des Solides, UMR 8502, CNRS,
Universit\'e Paris-Sud, F-91405 Orsay.}

\maketitle
\abstracts{ We study weak localization in chains of metallic rings. We show
  than nonlocality of quantum transport can drastically affect the behaviour
  of the harmonics of magnetoconductance oscillations. Two different
  geometries are considered~: the case of rings separated by long wires
  compared to the phase coherence length and the case of contacted rings. In
  a second part we discuss the role of decoherence due to electron-electron
  interaction in these two geometries. }

\section{Introduction}

At low temperature, quantum interferences of reversed electronic trajectories
are responsible for a small reduction of the averaged conductivity called the
{\it weak localization} (WL) correction. This correction is a manifestation of
quantum coherence which is always limited over a certain length scale, named
the {\it phase coherence length} $L_\varphi$. A way to extract this
important length scale in experiments is to use the magnetic field sensitivity
of the WL. For example the WL correction of an infinitely long wire of
rectangular section of width $W$ and area $S$ submitted to a perpendicular
magnetic field $\mathcal{B}$ is~\cite{AltAro81} 
$
\smean{\Delta\sigma} =-\frac{2e^2}{h}\,\frac1S\,
[\frac{1}{L_\varphi^2}+\frac13(\frac{e\mathcal{B}W}{\hbar})^2]^{-1/2} 
$ (in the following we will forget the $1/S$ factor). 
The width of the magnetoconductance (MC) curve provides a direct determination
of~$L_\varphi$.

\begin{figure}[!ht]
\centering
\includegraphics[scale=0.5]{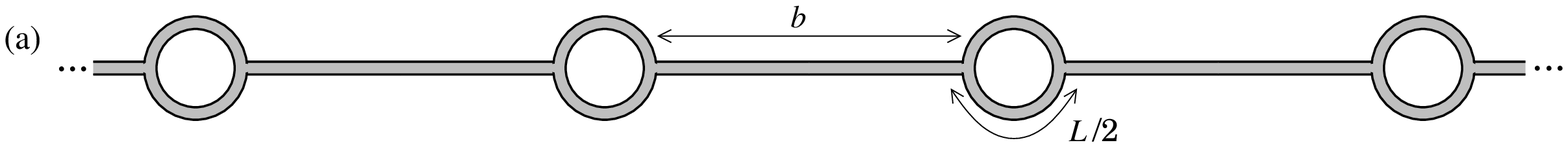}\\[0.25cm]
\includegraphics[scale=0.5]{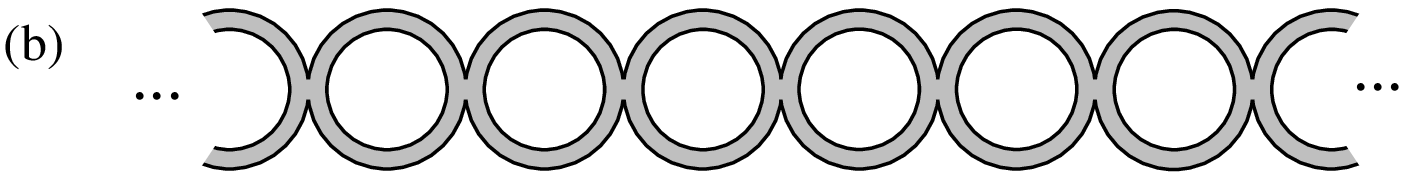}
\caption{\it Chains of rings. If we consider the regime $b\gg{L_\varphi}\gg L$ 
         the rings can be considered as independent in case (a) but not 
         in case (b).\label{fig:chains}}
\end{figure}

Another possibility to extract phase coherence length is to study arrays of
rings whose MC present oscillations as a function of the flux $\phi$ per ring
with period {\it half} the quantum flux $\phi_0=h/e$. These are the famous
Al'tshuler-Aronov-Spivak oscillations~\cite{AltAroSpi81} (AAS), observed in many
experiments~\cite{PanChaRamGan85,DolLicBis86,FerAngRowGueBouTexMonMai04,SchMalMaiTexMonSamBau07}.
In order to extract the phase coherence length from AAS oscillations, a
precise theoretical prediction for the behaviour of the AAS harmonics with the
phase coherence length is needed. Harmonics of the oscillations are defined as
$\smean{\Delta\sigma(\phi)}=\sum_{n}\Delta\sigma_n\EXP{4\pi\I{n}\phi/\phi_0}$.
A well-known expression has been derived in Ref.~\cite{AltAro81} for an
isolated ring of perimeter~$L$~:
\begin{equation}
  \label{AAS}
  \Delta\sigma_n=-\frac{2e^2}{h}L_\varphi\EXP{-|n|L/L_\varphi}
\end{equation}
However, in a real experiment where the ring is connected to wires, or
embedded in a larger network, this expression can only be relevant in the
regime $L_\varphi\ll{L}$. At the lowest temperatures, when
$L_\varphi\gtrsim{L}$, the AAS harmonics are strongly affected by the
surrounding wires since trajectories can expand outside the ring over
distances larger than the perimeter. It is the aim of this paper to discuss
the behaviour of AAS harmonics in chains of rings when $L_\varphi\gtrsim{L}$.
We will consider two cases represented on figure~\ref{fig:chains}~: in the
first situation the rings are separated by a distance $b\gg{L_\varphi}$ and
can therefore be considered as independent (however the connecting wires will
affect the AAS harmonics). In the second case rings are in contact and
harmonics can involve trajectories winding around several neighbouring rings.
In section~\ref{sec:ee}, we will see that when electron-electron interaction is 
the dominant process for decoherence, eq.~(\ref{AAS}) cannot be used even in 
the regime~$L\ll{L_\varphi}$.

\section{Nonlocality of quantum transport in chain of rings\label{sec:1}}

We consider an array of rings all pierced by the same flux $\phi$. The $n$-th
harmonic of the WL correction at a given point $x$ of a network can be
expressed as
\begin{equation}
  \label{harmWL}
  \Delta\sigma_n(x) = -\frac{2e^2D}{\pi}
  \int_0^\infty\D{t}\,\mathcal{P}_n(x,x;t)\,\EXP{-t/\tau_\varphi}
\end{equation}
where $D$ is the diffusion constant and $\tau_\varphi=L_\varphi^2/D$ the phase
coherence time. The factor $2$ stands for spin degeneracy.
$\mathcal{P}_n(x,x;t)$ is the probability that a particle diffusing into the
network comes back to its initial point $x$ in a time $t$, after having encircled a
flux $n\phi$. For example, in an isolated ring
$\mathcal{P}_n(x,x;t)=\frac1{\sqrt{4\pi{Dt}}}\EXP{-(nL)^2/4Dt}$ which
immediatly gives eq.~(\ref{AAS}). Except in translation invariant systems,
$\Delta\sigma_n(x)$ depends on $x$ and expression (\ref{harmWL}) must be
averaged over the network in a proper way described in Ref.~\cite{TexMon04}.

\vspace{0.15cm}

\noindent{\bf A ring with two arms.--}
The case of a ring connected to two arms has been studied in detail in
Ref.~\cite{TexMon05} where it has been shown that $ \mathcal{P}_n(x,x;t)
\simeq\frac{\sqrt{L/2}}{2(Dt)^{3/4}}\Psi(\frac{n\sqrt{2L}}{(Dt)^{1/4}}) $ for
time scales $t\gg{L}^2/D$ with $x$ inside the ring (the precise form of the
dimensionless function $\Psi(\xi)$ is inessential for the present discussion).
Compared with the isolated ring case, where the typical number of winding
scales with time as $n_t\sim{t}^{1/2}$, diffusion around the ring is slowed
down as $n_t\sim{t}^{1/4}$ due to the time spent in the arms. As a consequence
the harmonics of the conductance of a ring are given by~\cite{TexMon05}~:
$\Delta\sigma_n\propto{L_\varphi}^{3/2}\EXP{-|n|\sqrt{2L/L_\varphi}}$ (note
that the scaling $n\sim{L}_\varphi^{1/2}$ is analogous to the scaling of
winding with time $n_t\sim{t}^{1/4}$ since~$L_\varphi\sim{t}^{1/2}$).

\vspace{0.15cm}

\noindent{\bf The chain of distant rings.--}
The same argument holds for the chain of rings separated by a distance
$b\gg{L}_\varphi$ (figure~\ref{fig:chains}.a). In this case, averaging
properly $\Delta\sigma_n(x)$ inside the chain of $N_r$ rings, one finds that the
harmonics of the dimensionless conductance read~:
\begin{equation}
  \Delta{g}_n \simeq -
  \frac{N_r\,L^{1/2}{L_\varphi}^{3/2}}{\sqrt2\,[(N_r+1)b]^2}\,
  \EXP{-|n|\sqrt{2L/L_\varphi}}
  \hspace{0.5cm}\mbox{for}\hspace{0.5cm}
  b \gg L_\varphi \gg L
\end{equation}

\vspace{0.15cm}

\noindent{\bf The chain of attached rings.--}
If we now consider the network of figure~\ref{fig:chains}.b, we can show that
the probability reads~\cite{TexMon07}
$\mathcal{P}_n(x,x;t)\simeq\frac{L}{8\pi{Dt}}\EXP{-(nL)^2/4Dt}$ for
$t\gg{L}^2/D$. The AAS harmonics are given in this case by~\cite{TexMon07}~:
\begin{eqnarray}
  \Delta{g}_n   
  &\simeq& - \frac1{N_r}\,\frac2\pi\,
   \left[
     \ln(2L_\varphi/|n|L) + b_{n}  
   \right]
  \hspace{0.5cm}\mbox{ for }   L_\varphi/L \gg |n|  \\
  &\simeq& - \frac1{N_r}\,\sqrt{\frac2\pi}\,
  \frac{\EXP{-|n|\,L/L_\varphi}}{\sqrt{|n|\,L/L_\varphi}}
  \hspace{1.5cm}\mbox{ for }  |n|\gg L_\varphi/L\gg1  
\end{eqnarray}
where $b_n$ depends weakly on $n$ ($b_\infty=-\mathrm{C}$, the Euler constant).

\section{Decoherence due to electron-electron interaction\label{sec:ee}}

The above results rely on the fact that, in eq.~(\ref{harmWL}), the long times
have been cut off with an exponential damping $\EXP{-t/\tau_\varphi}$. However
it has been shown recently that this simple modelization does not account
correctly for the decoherence due to electron-electron interaction, which is
the dominant one at low temperature~\footnote{
  The exponential damping gives the correct shape of a MC of a 
  wire~\cite{PieGouAntPotEstBir03,AkkMon04} 
  with $L_\varphi\to\sqrt2L_N$ (see below for definition of $L_N$), 
  however this simple substitution gives 
  an incorrect result for AAS harmonics as explained below.
}.  
In this case, an alternative description was proposed by Al'tshuler, Aronov \&
Khmel'nitskii (AAK)~\cite{AltAroKhm82} but it is only recently that the
consequences for AAS oscillations have been
understood~\cite{LudMir04,TexMon05b,TexMon07b}. 

\vspace{0.15cm}

\noindent{\bf The model of AAK.--}
The length scale characterizing the efficiency of electron-electron
interaction to suppress phase coherence in wires is known as the {\it Nyquist
  length} $L_N=(\nu_0D^2/T)^{1/3}$ where $\nu_0$ is the density of states, $D$
the diffusion constant and $T$ the temperature ($\hbar=k_B=1$). In the model
of AAK, the random phase accumulated by an electron moving in the fluctuating
electric potential due to other electrons is included in the calculation of
the WL. The pair of reversed interfering trajectories picks a phase
$\EXP{\I\Phi[\mathcal{C}]}$, where $\mathcal{C}$ designates a closed diffusive
trajectory, and the harmonics of WL are given by
\begin{equation}
  \label{WLAAK}
  \Delta\sigma_n \sim -\sum_{\mathcal{C}_n}
  \smean{ \EXP{\I\Phi[\mathcal{C}_n]} }_V
  =-\sum_{\mathcal{C}_n}\EXP{-\frac12\smean{ \Phi[\mathcal{C}_n]^2 }_V}
\end{equation}
The sum runs over all closed trajectories with winding~$n$ (a proper
 formulation of eq.~(\ref{WLAAK}) requires a path integral).
Gaussian fluctuations of the electric potential are given by the
fluctuation-dissipation theorem 
$
\smean{V(\vec{r},t)V(\vec{r}\,',t')}_V=
\frac{2e^2}{\sigma_0}T\delta(t-t')P_d(\vec{r},\vec{r}\,')
$
(written here in the classical limit $T\ll\omega$), 
where $\sigma_0$ is the classical Drude conductivity. $P_d$ is solution
of the diffusion equation
$-\Delta{P}_d(\vec{r},\vec{r}\,')=\delta(\vec{r}-\vec{r}\,')$ and therefore
depends on the topology of the system. Then~\cite{TexMon05b}
\begin{equation}
  \smean{ \Phi[x(\tau)]^2 }_V
  = \frac{4D}{L_N^3}
  \int_0^t\D\tau\,
  \left[
    P_d(x(\tau),x(\tau)) - P_d(x(\tau),x(t-\tau))
  \right]
\end{equation}
where $\mathcal{C}\equiv(x(\tau),\,0\leq\tau\leq{t}\,|\,x(0)=x(t))$ is a
closed diffusive path. The crucial point is that the simple exponential
damping of eq.~(\ref{harmWL}) is replaced in eq.~(\ref{WLAAK}) by a functional
of the trajectory $\EXP{-\frac12\smean{\Phi[\mathcal{C}_n]^2}_V}$. Therefore 
decoherence is now network-dependent and {\it a priori} sensitive to the nature of
trajectories (in particular whether they do enclose a magnetic flux or not).

\vspace{0.15cm}

\mathversion{bold}
\noindent{\bf The limit $L_N\ll{L}$.--}
\mathversion{normal} The model described above was applied to the case of a
single ring~\cite{LudMir04,TexMon05b,TexMon07b}. 
The result for an
isolated ring is relevant to describe arrays of rings in the limit $L_N\ll{L}$
where winding trajectories hardly exit from a ring, which makes rings
independent from each other. For the chain of distant rings
(figure~\ref{fig:chains}.a) we have
\begin{equation}
  \label{LMTM}
  \Delta{g}_n \sim - \frac{N_rL}{[(N_r+1)b]^2} L_N  \, 
  \EXP{-|n|\frac\pi8(L/L_N)^{3/2}} 
  \sim \frac{\EXP{-n\,L^{3/2}T^{1/2}} }{T^{1/3}}
  \hspace{0.5cm} \mbox{for }  L_N \ll L \ll b
\end{equation}
(for the case of the chain of rings in contact (figure~\ref{fig:chains}.b),
$(N_r+1)b$ in the denominator is replaced by~$N_rL/4$). Whereas the time
characterizing efficiency of electron-electron interaction to suppress phase
coherence in a wire is the Nyquist time $\tau_N=L_N^2/D\propto{T}^{-2/3}$, it
was shown in Refs.~\cite{LudMir04,TexMon05b} that the behaviour (\ref{LMTM})
is related to a new time scale characterizing decoherence for winding
trajectories~: $\tau_c=\tau_N^{3/2}/\tau_L^{1/2}\propto{T}^{-1}$, where
$\tau_L=L^2/D$ is the Thouless time of the ring.

\vspace{0.15cm}

\noindent{\bf The chain of distant rings.--}
If we consider a ring connected to long arms, winding trajectories spend most
of the time in the arms~\cite{TexMon05} and the decoherence mostly occurs in
the arms. Therefore decoherence occurs on a time scale $\tau_N$, like in a
wire. The function $\smean{\EXP{\I\Phi[\mathcal{C}]}}_V$ for a wire was
studied in Ref.~\cite{MonAkk05}. Using this result and the winding properties
recalled in section~\ref{sec:1} leads to~\cite{TexMon05b} 
\begin{eqnarray}
  \Delta{g}_n &\sim& -\frac{N_rL^{1/2}L_N^{3/2}}{[(N_r+1)b]^2}
  \hspace{7cm} \mbox{for }  n^2\ll L_N/L \\
  &\sim& -\frac{N_rL^{1/2}L_N^{3/2}}{[(N_r+1)b]^2}
  \left(\frac{n^2L}{L_N}\right)^{7/12}
  \!\!\!\EXP{-\kappa_2|n|\sqrt{L/L_N}}
  \sim \frac{\EXP{ -n\,L^{1/2}T^{1/6} }}{T^{11/36}}
  \hspace{0.5cm} \mbox{for }  n^2\gg L_N/L \:,
\end{eqnarray}
where~$\kappa_2=\sqrt2|u_1|^{1/4}\simeq1.421$.

\vspace{0.15cm}

\noindent{\bf The chain of attached rings.--}
In this case, the nature of decoherence was shown to be closely related to the
one of a wire since diffusion along the chain is reminiscent of a 1d
diffusion and again occurs on time scale $\tau_N$~\cite{TexMon07}~:
\begin{eqnarray}
  \Delta{g}_n
  &\simeq& -\frac{2}{N_r\pi}\,\ln(L_N/|n|L)+{\rm cste}
  \hspace{2cm} \mbox{for }  |n|\ll L_N/L  \\ 
  &\simeq& -\frac{1}{N_r|u_1|^{3/2}}\, \EXP{ -\kappa_3|n|{L}/{L_N} }
  \sim \EXP{ -nL T^{1/3} }
  \hspace{0.5cm} \mbox{for }  |n| \gg L_N/L 
\end{eqnarray}
where $\kappa_3=2^{-1/3}|u_1|^{1/2}\simeq0.801$.

\section{Conclusion}

We have considered networks of connected rings, made of weakly disordered
wires. We have first shown that geometrical effects can strongly modify the
exponential behaviour of AAS harmonics well-known for an isolated ring, since
trajectories can now explore the network around each ring. In the second part
we have shown that decoherence due to electron-electron interaction is
sensitive to geometry, a second reason that modifies the simple AAS result. An
interesting experiment would be to compare precisely AAS oscillations for the
two networks of figure~\ref{fig:chains} in the low temperature
regime~$L_N\gg{L}$.



\section*{References}

\bibliographystyle{phreport}

\begin{thebibliography}{100}

\bibitem{AltAro81}
B.~L. Al'tshuler and A.~G. Aronov,
 JETP Lett. {\bf 33}(10), 499 (1981).

\bibitem{AltAroSpi81}
B.~L. Al'tshuler, A.~G. Aronov, and B.~Z. Spivak,
 JETP Lett. {\bf 33}(2), 94 (1981).

\bibitem{PanChaRamGan85}
B.~Pannetier, J.~Chaussy, R.~Rammal, and P.~Gandit,
 Phys. Rev.~B {\bf 31}(5), 3209 (1985).

\bibitem{DolLicBis86}
G.~J. Dolan, J.~C. Licini, and D.~J. Bishop,
 Phys. Rev. Lett. {\bf 56}(14), 1493 (1986).

\bibitem{FerAngRowGueBouTexMonMai04}
M.~Ferrier, L.~Angers, A.~C.~H. Rowe, S.~Gu{\'e}ron, H.~Bouchiat, C.~Texier,
  G.~Montambaux, and D.~Mailly,
 Phys. Rev. Lett. {\bf 93}, 246804 (2004).

\bibitem{SchMalMaiTexMonSamBau07}
F.~Schopfer, F.~Mallet, D.~Mailly, C.~Texier, G.~Montambaux, L.~Saminadayar,
  and C.~B\"auerle,
 Phys. Rev. Lett. {\bf 98}, 026807 (2007).

\bibitem{TexMon04}
C.~Texier and G.~Montambaux,
 Phys. Rev. Lett. {\bf 92}, 186801 (2004).

\bibitem{TexMon05}
C.~Texier and G.~Montambaux,
 J.~Phys.~A: Math. Gen. {\bf 38}, 3455--3471 (2005).

\bibitem{TexMon07}
C.~Texier and G.~Montambaux,
 in preparation  (2007).

\bibitem{PieGouAntPotEstBir03}
F.~Pierre, A.~B. Gougam, A.~Anthore, H.~Pothier, D.~Esteve, and N.~O. Birge,
 Phys. Rev.~B {\bf 68}, 085413 (2003).

\bibitem{AkkMon04}
{\'E}.~Akkermans and G.~Montambaux,
 {\it Phy\-si\-que m\'esos\-co\-pi\-que des \'elec\-trons et des
  pho\-tons},
 EDP Sciences, CNRS \'editions, 2004.
{\it Mesoscopic physics of electrons and photons},
  Cambridge University Press, 2007.

\bibitem{AltAroKhm82}
B.~L. Altshuler, A.~G. Aronov, and D.~E. Khmelnitsky,
 J.~Phys.~C: Solid St. Phys. {\bf 15}, 7367 (1982).

\bibitem{LudMir04}
T.~Ludwig and A.~D. Mirlin,
 Phys. Rev.~B {\bf 69}, 193306 (2004).

\bibitem{TexMon05b}
C.~Texier and G.~Montambaux,
 Phys. Rev.~B {\bf 72}, 115327 (2005)~;
%
{\it ibid} {\bf 74}, 209902(E) (2006).

\bibitem{TexMon07b}
C.~Texier and G.~Montambaux,
 Comment on 
 Ref.~\cite{LudMir04},
 submitted  (2007).

\bibitem{MonAkk05}
G.~Montambaux and E.~Akkermans,
 Phys. Rev. Lett. {\bf 95}, 016403 (2005).

\end{thebibliography}

\end{document}